\documentclass[11pt,twoside]{article}
\usepackage{asp2014}

\aspSuppressVolSlug
\resetcounters

\begin{document}

\title{Improving Software Citation and Credit}
\author{Alice Allen$^1$, G. Bruce Berriman$^2$, Kimberly DuPrie$^3$, Jessica Mink$^4$, Robert Nemiroff$^5$, Thomas Robitaille$^6$, Lior Shamir$^7$, Keith Shortridge$^8$, Mark Taylor$^9$, Peter Teuben$^{10}$, and John Wallin$^{11}$
\affil{$^1$Astrophysics Source Code Library}
\affil{$^2$Infrared Processing and Analysis Center, California Institute of Technology}
\affil{$^3$Space Telescope Science Institute}
\affil{$^4$Harvard-Smithsonian Center for Astrophysics}
\affil{$^5$Michigan Technological University}
\affil{$^6$Max Planck Institute for Astronomy}
\affil{$^7$Lawrence Technological University}
\affil{$^8$Australian Astronomical Observatory}
\affil{$^9$University of Bristol}
\affil{$^{10}$University of Maryland}
\affil{$^{11}$Middle Tennessee State University}}

\paperauthor{Alice Allen}{aallen@ascl.net}{}{Astrophysics Source Code Library}{}{}{}{}{}
\paperauthor{G. Bruce Berriman}{gbb@ipac.caltech.edu}{orcid.org/0000-0001-8388-534X}{California Institute of Technology}{IPAC}{Pasadena}{CA}{91125}{USA}
\paperauthor{Kimberly DuPrie}{kduprie@stsci.edu}{}{Space Telescope Science Institute}{}{}{}{}{}
\paperauthor{Jessica Mink}{jmink@cfa.harvard.edu}{orcid.org/0000-0003-3594-1823}{Harvard-Smithsonian Center for Astrophysics}{}{}{}{}{}
\paperauthor{Robert Nemiroff}{nemiroff@mtu.edu}{orcid.org/0000-0002-4505-6599}{Michigan Technological University}{}{}{}{}{}\paperauthor{Thomas Robitaille}{robitaille@mpia.de}{orcid.org/0000-0002-8642-1329}{Max Planck Institute for Astronomy}{ }{Heidelberg}{ }{D-69117}{Germany}
\paperauthor{Lior Shamir}{lshamir@ltu.edu}{orcid.org/0000-0002-6207-1491}{Lawrence Technological University}{}{}{}{}{}\paperauthor{Keith Shortridge}{ks@aao.gov.au}{}{Australian Astronomical Observatory}{}{}{}{}{}
\paperauthor{Mark Taylor}{M.B.Taylor@bristol.ac.uk}{}{University of Bristol}{}{}{}{}{}
\paperauthor{Peter Teuben}{teuben@astro.umd.edu}{orcid.org/0000-0003-1774-3436}{University of Maryland}{}{}{}{}{}
\paperauthor{John Wallin}{John.Wallin@mtsu.edu}{orcid.org/0000-0001-5678-8325}{Middle Tennessee State University}{}{}{}{}{}
\begin{abstract}

The past year has seen movement on several fronts for improving software citation, including the Center for Open Science's Transparency and Openness Promotion (TOP) Guidelines, the Software Publishing Special Interest Group that was started at January's AAS meeting in Seattle at the request of that organization's Working Group on Astronomical Software, a Sloan-sponsored meeting at GitHub in San Francisco to begin work on a cohesive research software citation-enabling platform, the work of Force11 to ``transform and improve'' research communication, and WSSSPE's ongoing efforts that include software publication, citation, credit, and sustainability.

Brief reports on these efforts were shared at the BoF, after which participants discussed ideas for improving software citation, generating a list of recommendations to the community of software authors, journal publishers, ADS, and research authors. The discussion, recommendations, and feedback will help form recommendations for software citation to those publishers represented in the Software Publishing Special Interest Group and the broader community.

\end{abstract}

\section{Introduction}

Providing credit to code authors through citation has been a recurring topic in previous Birds of a Feather (BoF) sessions sponsored by the Astrophysics Source Code Library (ASCL)\footnote{\url{http://ascl.net/}} at ADASS meetings: (\textit{Bring out your codes! Bring out your codes!} \citep{2013ASPC..475..383A} and \textit{Ideas for advancing code sharing} \citep{2014ASPC..485....3T}. This BoF continued the work started at previous BoFs on that topic, and represents a topic being addressed by the ASCL at astronomy software sessions and topical meetings.

The BoF opened with a short presentation by Bruce Berriman and Alice Allen. Berriman described a Software Publishing Special Interest Group (SPSIG) meeting held by the ASCL at the American Astronomical Society (AAS) meeting in January 2015 to discuss software citation; the SPSIG was formed at the request of the AAS's Working Group on Astronomical Software (WGAS). The meeting was attended by publishers and editors from AAS journals, Springer, IOP, Cambridge University Press and Oxford University Press, software authors, representatives from the Astrophysics Data System (ADS)\footnote{\url{http://adswww.harvard.edu/}}, GitHub and projects such as the Large Synoptic Survey Telescope (LSST), researchers, and others. Berriman summarized a working Google document\footnote{\url{http://tinyurl.com/nqtf29h}} that captured the deliberations at that meeting (essentially, a summary of the current state of software citation in astronomy). He also presented opinions on what constitutes a citable work, the difference between attribution and citation, and a restatement of the distinction between citation and attribution by Christine Borgman. 

Allen reported on recent efforts by software citation workgroups formed at the 3rd Workshop on Sustainable Software for Science: Practice and Experiences (WSSSPE3)\footnote{\url{http://wssspe.researchcomputing.org.uk/wssspe3/}} and Force11;\footnote{\url{https://www.force11.org/}} as these efforts are very similar, the WSSSPE group has now joined the Force11 efforts. She also reported on the Center for Open Science's Transparency and Openness Promotion (TOP) Guidelines\footnote{\url{https://osf.io/9f6gx/}} and a Sloan-sponsored meeting at GitHub in San Francisco to begin work on a cohesive research software citation-enabling platform. The slides from Berriman's and Allen's presentation are available online,\footnote{\url{https://docs.google.com/presentation/d/1fLaWPsCWgVmGqO8mhKBWK6dvV7VpQy9MtubtL2zAEzQ/edit?usp=sharing}} as are other resources and links.\footnote{\url{http://ascl.net/wordpress/?p=1532}}

\section{Group Discussion}

The very lively discussion among the 40 attendees was moderated by Keith Shortridge; a Google document\footnote{\url{http://tinyurl.com/o62gxlk}} captured some of the discussion and was later shared and augmented by some of the attendees. Different citation methods mentioned in Berriman's presentation were discussed; a software description paper has been the most common way to cite software that has been used in a research project. Even with a software description paper available to use for citation, many codes used in research do not receive a formal citation in research papers. Alberto Accomazzi pointed this out by pulling some quick statistics from  ADS for the {\tt DAOPHOT} package, and reported that the {\tt DAOPHOT} code description paper has over 3,000 citations to it, yet the software is mentioned in more than 6,000 papers. Accomazzi supplied more exact numbers after the meeting: as of November 11, the {\tt DAOPHOT} code description paper had 4,035 formal citations and the software was mentioned in 3,061 papers that did not cite it formally. It has been previously noted that quantitative measures of the impact of software on the astronomy community are hard to derive in the absence of a culture of citation: e.g., ``... although some 22,000 peer-reviewed papers mention the VLA radio telescope, only 68 formally acknowledge the use of AIPS and only 59 acknowledge use of CASA, the two dominant reduction and analysis packages for radio interferometry data.''\citep{2015A&C....11..190H}

A recent experiment among some journals to request code with papers, requiring a code author to provide a tarball of the software and turn over copyright of it to a publisher, was discussed with vigor. This practice did not receive any support among those assembled, and later (and ongoing) discussion made clear how concerned software authors are about this path. Indeed, this practice was unanimously condemned.

The need to make a distinction between publishing software and making it available -- releasing it -- was discussed and then a request for the group to start focusing on possible recommendations and actionable suggestions was made and followed.

\section{Ideas from the Collected Masses}
Some of the suggestions made for improving software citation and credit were directed to specific parts of the community to do or to use, whereas others were more general or assumed to be for the ASCL or other entities involved in software, and included: 

\begin{itemize}
  \item For authors: Provide information as to what software should be cited. Cite the first-level software; a manuscript author is not responsible for citing software dependencies unless there are specific instructions from first-level software author for citing them.
  \item For authors: Do not cite GitHub directly. Use Internet Archive, ASCL, Zenodo, Figshare, etc.
  \item For publishers: Do not count references against the word count.
  \item For ADS: Include software in categorization of entries.
   \item For the community: Encourage your university to ask about software on the annual research activity report.
  \item For the community: Write a wiki article for AstroBetter and the AAS newsletter and other places on how to release software for citation, and how to cite software. 
  \item For the community: Create and award a prize for software contributions.
  \item For the community: Create a video on how to release and cite software effectively.
  \item For the community: Collect and publish stories from people who have released their software and what their views are on releasing software.
\end{itemize}

\section{Conclusions}

Clearly there is a role for each person in the community to contribute to the goal of improving software citation and credit. Software authors can release their codes, follow one of several paths for making their code easily citable, and specify clearly and obviously how they want their software cited. Researchers using software in their work can cite computational methods as their authors specify, and journal editors can insist that codes be cited properly in the manuscripts they accept. Publishers can require software citations that are properly formatted so indexers can pick up and track the citations, and can remove length restrictions that prohibit methods citations. Those serving the community, such as ADS and ASCL, can promote better software citation by sharing information about citations and encouraging the community to improve, and individuals can push for inclusion of software activities in consideration for promotions and tenure. 

\acknowledgements

Our thanks to all the attendees for their comments, ideas, and advocacy, to the ADASS POC for accepting our BoF proposal, and to Keith Shortridge for moderating the spirited discussion.

\bibliography{B4}

\begin{thebibliography}{}
\expandafter\ifx\csname natexlab\endcsname\relax\def\natexlab#1{#1}\fi
\expandafter\ifx\csname url\endcsname\relax
  \def\url#1{\texttt{#1}}\fi
\expandafter\ifx\csname urlprefix\endcsname\relax\def\urlprefix{URL }\fi
\providecommand{\eprint}[2][]{\url{#2}}

\bibitem[{{Allen} et~al.(2013){Allen}, {Berriman}, {Brunner}, {Burger},
  {DuPrie}, {Hanisch}, {Mann}, {Mink}, {Sandin}, {Shortridge}, \&
  {Teuben}}]{2013ASPC..475..383A}
{Allen}, A., {Berriman}, B., {Brunner}, R., {Burger}, D., {DuPrie}, K.,
  {Hanisch}, R.~J., {Mann}, R., {Mink}, J., {Sandin}, C., {Shortridge}, K., \&
  {Teuben}, P. 2013, in Astronomical Data Analysis Software and Systems XXII,
  edited by D.~N. {Friedel}, vol. 475 of Astronomical Society of the Pacific
  Conference Series, 383. \eprint{1212.1915}

\bibitem[{{Hanisch} et~al.(2015){Hanisch}, {Berriman}, {Lazio}, {Emery},
  {Evans}, {McGlynn}, \& {Plante}}]{2015A&C....11..190H}
{Hanisch}, R.~J., {Berriman}, G.~B., {Lazio}, T. J.~W., {Emery}, S., {Evans},
  J.~., {McGlynn}, T.~A., \& {Plante}, R.~A. 2015, Astronomy and Computing, 11,
  190

\bibitem[{{Teuben} et~al.(2014){Teuben}, {Allen}, {Berriman}, {DuPrie},
  {Hanisch}, {Mink}, {Nemiroff}, {Shamir}, {Shortridge}, {Taylor}, \&
  {Wallin}}]{2014ASPC..485....3T}
{Teuben}, P., {Allen}, A., {Berriman}, B., {DuPrie}, K., {Hanisch}, R.~J.,
  {Mink}, J., {Nemiroff}, R.~J., {Shamir}, L., {Shortridge}, K., {Taylor},
  M.~B., \& {Wallin}, J.~F. 2014, in Astronomical Data Analysis Software and
  Systems XXIII, edited by N.~{Manset}, \& P.~{Forshay}, vol. 485 of
  Astronomical Society of the Pacific Conference Series, 3. \eprint{1312.7352}

\end{thebibliography}

\end{document}